# STELLAR MULTIPLICTY IN THE σ ORIONIS CLUSTER: A REVIEW

*By José A. Caballero*
*Centro de Astrobiología (CSIC-INTA), Madrid, Spain*

The nearby, young, extinction-free σ Orionis cluster is acknowledged as a cornerstone for studying the stellar and substellar formation and evolution, especially at very low masses. However, the relatively limited knowledge of multiplicity in the cluster is dispersed among numerous papers and proceedings with very different aims and, therefore, is difficult to apprehend even for researchers in the subject. Here, I review comprehensively the sexdecuple system σ Ori in the central arcminute, and all close ($\rho$ = 0.4–4 arcsec) and wide ($\rho$ > 4 arcsec) binaries, triples, hierarchical multiples, and spectroscopic binaries in the cluster that have been reported in the literature until 2014. I conclude with a brief enumeratation of the next steps that should be taken in the near future for having a clear picture of the multiplicity in the σ Orionis cluster.

*Introduction*

Regardless of what we call them *The Big Three*, *Ζωνη* (*Zone*), *Balteus*, *Shen*, *Tres Marías*, *Oriongürtel, Ceinture d'Orion, Jacob's Rod*, *Magi, Golden Yard-arm, Işus Trikāṇdā*[1] or the Orion's Belt, the three blue supergiant stars Alnitak (ζ Ori), Alnilam (ε Ori) and Mintaka (δ Ori) form the most prominent asterism in the sky. Since the ancient times, they have attracted the attention of sky watchers, seamen, astronomers, and even poets. The fourth brightest star in the Orion's Belt, just one degree south of Alnitak and about 2 mag fainter than the three main stars, is σ Ori (*Quae ultimam baltei præcedit ad austrum*, 48 Ori, HD 37468). In spite of not having been catalogued by Ptolemaios (137 AD) or Ulugh Beg (1437)[2], σ Ori is, with $V \sim 3.8$ mag, visible with the naked eye as a single star[3,4,5]. However, it is actually a Trapezium-like system that contains *six* early-type stars, including the massive astrometric "binary" σ Ori AB and the rotationally variable B2Vp star σ Ori E. Remarkably, σ Ori is the principal source of IC 434, the emission nebula against which the famous Horsehead dark nebula (Barnard 33) is silhouetted.

The σ Ori multiple system has taken a great significance in the last two decades because of the homonymous star cluster that surrounds it, σ Orionis[6,7,8], which has become one of the best laboratories in the sky to study star and brown dwarf formation[9,10,11]. With an age of 3 Myr, and a distance 385 pc, nearly free of extinction, the open cluster in the Ori OB1b association has been subject of numerous works on X-ray emission[12,13], disc frequency[14,15,16], presence of Herbig-Haro objects[17,18], photometric variability[19,20,21,22], accretion rates and frequency[23,24,25,26] and, especially, mass function at low stellar and substellar masses, down to below the deuterium burning limit[27,28,29,30,31]. In spite of its importance as a test-bed for models of high- and low-mass star formation and evolution, the issue of the multiplicity in the σ Orionis cluster has not bet addressed yet in an exhaustive manner. Here I compile *all* available information on multiplicity in σ Orionis prior to 2014. As with any other review, retrieved and re-measured data are so heterogenous that can hardly be the subject of a serious statistical analysis, but instead shed light on previous scatterred, difficult-to-find works on the topic, as well as pave the way for future high-resolution imaging and spectroscopic surveys.

*A bit of history*

The first multiple system in σ Orionis was reported in 1779 in the *Tabula Nova Stellarum Duplicium* by Mayer[32], who catalogued what we call now σ Ori AB, D, and E[33]. A few weeks after the announcement, W. Herschel promptly made the first accurate measures of angular separation ($\rho$) and position angle ($\theta$) of the triple system[34,35]. Decades later, Dawes[36] discovered σ Ori C, five magnitudes fainter than AB. It was not until the end of the 19th century when Burnham[37] realized that the brightest star in the (up-to-then) quadruple σ Ori system was in its turn a very close binary of $\rho \sim 0.25$ arcsec, later dubbed "the most massive binary with an astrometric orbit"[38,39]. A sixth, massive, OB-type component, which made Burnham's pair a triple, was confirmed recently with high-resolution spectroscopy by Simón-Díaz *et al.*[40]. As described below, the existence of new low-mass stars at less than 1 arcmin to σ Ori AB may make the system to be at least a sexdecuple[41,42].

The Washington Double Star catalogue (WDS[43]) lists 13 cluster pairs with angular separations between 3.3 arcsec and 8.7 arcmin (Table I). Eight pairs are associated to σ Ori itself, and the remaining five ones to two sky asterisms containing the stars HD 294271/2 and TX Ori/TY Ori. Tabulated astrometric measurements of $\rho$ and $\theta$ come from WDS except for σ Ori A,B (average of three measurements by Docobo *et al.*[44] in 2004.10), AB,C (by 2MASS[45] in 1998.828), AB,D (by *Hipparcos*[46] in 1991.25), AB,E (by Tycho-2[47] in 1991.25), and AB,G (by Caballero[48] in 2003.688). Compare these values with those provided by Mason *et al.*[49]. The notations σ Ori 'F' through 'I' are not used at all in the literature. Fig. 1 illustrates the heart of the σ Ori trapezium.

Note that the WDS codes in the third column do not match the discoverer in all cases, being Mayer and Dawes, and not Struve, the discoverers of σ Ori D and E, and C, respectively. Besides, σ Ori IRS1, which is described extensively below, was *not* discovered by Turner *et al.*[52], not did they provided astrometric data for the first time.

As a final comment on Table I, the third star in the WDS 05386–0244 system, [WB2004] 21[56], is a cluster interloper. There is a fourth star near WDS 05386–0244, namely Mayrit 508194[57,58] (S Ori J053836.7–024414), which is not catalogued by WDS. It is however a bona fide cluster member located at $\rho \sim 45$ arcsec, $\theta \sim 89$ deg to the primary and only $\rho \sim 25$ arcsec, $\theta \sim 152$ deg to the secondary.

Table I

*Wide pairs in the σ Orionis cluster tabulated by the Washington Double Star catalogue*

| WDS | Primary (Mayrit No.) | Code | Secondary (Mayrit No.) | ρ [arcsec] | θ [deg] |
|---|---|---|---|---|---|
| 05387–0236 | σ Ori Aa, Ab (Aa, Ab) | BU 1032 AB[37] | σ Ori B (B) | 0.250±0.006 | 100.9±0.5 |
| | | STF 762 AB,C[50,51] | σ Ori C (11238) | 11.43±0.06 | 237.8±0.4 |
| | | STF 762 AB,D[50,51] | σ Ori D (13084) | 12.98±0.03 | 84.20±0.12 |
| | | STF 762 AB,E[50,51] | σ Ori E (42062) | 41.555±0.004 | 61.720±0.008 |
| | | STF 3135 AB,F[50,51] | HD 294271 (208324) | 209.6 | 324 |
| | | TRN 19 AB,G[52] | σ Ori IRS1 (3020) | 3.32±0.06 | 19.6±1.4 |
| | | SHJ 65 AB,H[53] | HD 37525 (306125) | 306.7 | 125 |
| | | SHJ 65 AB,I[53] | HD 37564 (524060) | 524.1 | 60 |
| 05386–0233 | HD 294271 (208324) | STF 761 AB[50,51] | HD 294272 A (182305) | 67.8 | 203 |
| | | STF 761 AC[50,51] | HD 294272 B (189303) | 71.8 | 209 |
| | | STF 761 AD[50,51] | BD–02 1324D (240322) | 32.9 | 309 |
| 05386–0244 | TX Ori (521199) | PLT 1 AB[54] | TY Ori (489196) | 40.0 | 55 |
| | | DAM 51 AC[55] | [WB2004] 21 (...) | 16.9 | 33 |

*The σ Ori trapezium*

"*Perhaps the finest multiple star in the sky visible to both northern and southern observers, σ Ori is a system of five stars, four of which are visible in a small telescope [...]. This region of Orion's belt is an astrophotographer's dream*"[59]. These stargazer's sentences illustrate not only the great interest of the σ Ori trapezium to amateur astronomers, but also an extended misconception about the actual number of early-type stars in the system. Regrettably, this error is also widespread among professional astronomers, who besides stress on the importance of the late-O and early-B stars in σ Ori as a key trigger for the stellar formation in the Horsehead Nebula[60,61,62,63].

The astrophysical parameters of all the stars in the central cusp of the σ Orionis cluster are important for investigating a wealth of topics: (*i*) the high-mass stars in the σ Ori trapezium shape many, if not all, photo-dissociation regions, remnant molecular clouds, bright-rimmed clouds, cometary globules, and small reflection clouds in the Ori OB1b sub-association, especially the Horsehead Nebula[64,65,66,67] and a nearby bow-shock dust wave[68,69]; (*ii*) turbulence injection in the intracluster medium via ultraviolet heating and photo-erosion of pre-existing cores, also originated in high-mass stars, may explain the high frequency of brown dwarfs and

'isolated planetary-mass objects' in the cluster[70,71,72,73,74]; (*iii*) the top of the (initial) mass function, up to ~20 $M_\odot$[58,75,76]; (*iv*) heliocentric distances and dynamical masses of the most massive components[39]; and, of course, (*v*) multiplicity at all mass domains.

Table II
*Cluster members not listed in WDS and at $\rho$ < 1.0 arcmin to σ Ori A*

| Mayrit No. | Sp. type | $\rho$ [arcsec] | $\theta$ [deg] | Disc | X-rays | Li I | Hα |
|---|---|---|---|---|---|---|---|
| 21023[77,78] | M:[78] | 20.61±0.10 | 23.5±0.4 | ... | Yes[13,78] | ... | ... |
| 30241[78] | K-M:[78] | 29.81±0.06 | 241.14±0.16 | Yes[78,79] | Yes[78,80] | ... | ... |
| 36263[78] | M3.5:[26] | 35.73±0.06 | 263.17±0.11 | Yes[79] | ... | Yes[26] | Yes[26] |
| 50279[78] | M6.0:[26] | 50.07±0.06 | 279.06±0.06 | Yes[79] | No?[13,78] | Yes[26] | Yes[26] |
| 53049[58] | M1.0:[26] | 53.47±0.06 | 49.17±0.09 | Yes[78,79] | Yes[12] | Yes[26] | Yes[26] |
| 53144[78] | M5.0:[26] | 53.36±0.06 | 144.18±0.10 | ... | Yes[13,80] | Yes[26] | Yes[26] |

Apart from σ Ori Aa, Ab, B, C, D, E and IRS1 in Table I, there are another six late-type cluster members at $\rho$ < 1.0 arcmin to σ Ori A, with features of youth (near-infrared excess due to circumstellar disc, X-ray emission, Li I $\lambda$6707.8 nm in absorption, Hα $\lambda$656.3 nm in emission) and found in all-sky surveys, which are listed in Table II. Tabulated angular separations and position angles were measured from 2MASS[45] data obtained on 1998.828. There are, however, more faint objects in the central arcminute of the cluster, generally ascribed as components of the σ Ori multiple system. Below, I enumerate and review the 16 known stars in the 1-arcmin radius central area. In parenthesis, *N* indicates the increasing number of components. For this enumeration and in the rest of the paper, I use the Mayrit nomenclature. The Mayrit catalogue[11] and its subsequent additions and corrections is up to now the most comprehensive list of σ Orionis members and candidates free of contamination by fore- and background interlopers, and is rutinely used by other authors worldwide. Given the diversity of surveys and searches in the open cluster, the homogeneous use of the Mayrit nomenclature facilitates the easy recognition and writing of most objects in σ Orionis, which typical names are as complex as [BZR2001] S Ori J053825.4–024241, [W96] rJ053827–0242, 2MASS J05382732–0243247 or [HHM2007] 488, just to put a few examples. Besides, the Mayrit number provides the angular separation and position angle with respect to the cluster centre (e.g., Mayrit 42062 is the star at $\rho$ = 42 arcsec and $\theta$ = 62 deg with respect to σ Ori AB).

- Mayrit Aa,Ab,B = σ Ori Aa,Ab,B (*N* = 3; see Table I). Since its discovery by Burnham[37], the tight binary BU 1032 AB in the very centre of the cluster has not completed a whole revolution yet. Fortuitously, the first astrometric orbit of σ Ori AB was computed by Siegrist[81]. Her four-page manuscript was published in Spanish by a journal with a very limited distribution, and it may have gone unnoticed forever if not for colleagues at the Universidad Complutense de Madrid. As she remarked, given the low quality of her data (and small and spare time coverage – the latest astrometric observation was done by Müller in 1950.20; two world wars happened in between), her orbit should be considered provisional. In spite of it, she already pointed at the pair having a low eccentricity and a period longer than a century. Better orbital parameters were determined later by other authors[82,83,84,85]. To date, the most recent astrometric orbit was presented by Turner *et al.*[86], who tabulated $P$ = 156.7±3.0 yr, $a$ = 0.2662±0.0021 arcsec, and $e$ = 0.0515±0.0080. Although suspected by some authors during the whole 20th century[87,88], it was not until 120 years after the Burnham's discovery that Simón-Díaz *et al.*[40] demonstrated the binary to be actually triple. They first obtained solutions for the radial-velocity curves of components Aa and Ab, resulting in a highly eccentric orbit ($e \sim 0.78$) with a spectroscopic period of 143.5±0.5 d (400 times smaller than the astrometric period). They assigned spectral types O9.5 V, B0.5 V, and early-B to Aa, Ab, and B, respectively (before that, it was though that σ Ori B was B0.5 V). Lately, two teams have been able to resolve σ Ori Aa and Ab, which are separated by less than 8 mas, with the Navy Precision Optical Interferometer[89] and Michigan Infra-Red Combiner for the CHARA Interferometer[90]. Using unpublished radial-velocity data, both teams provided preliminary estimations of stellar masses ($M_{Aa}$ = 15.6–16.7 $M_\odot$ and $M_{Ab}$ = 12.4–12.8 $M_\odot$) and distance ($d$ = 380–385 pc, which matches Caballero's[39] triple-scenario distance of $d \approx 385$ pc). Simón-Díaz *et al.* (in prep.) provide a detailed spectroscopic analysis of the triple system and calculate an improved period of the spectroscopic-interferometric pair with an accuracy of only 11 min.

- Mayrit 3020 A,B = σ Ori IRS1A,B ($N$ = 5; see Table I). A "dust cloud next to σ Ori" was discovered in the mid-infrared by van Loon & Oliveira[91]. They explained the dense core and extended emission in a fan-shaped morphology, pointed away from the high-mass stellar system, as a photo-evaporating proto-planetary disc. Next, Caballero[92] imaged the central star of σ Ori IRS1 for the first time and, later, again Caballero[48] reported the first astrometric observation with near-infrared adaptive optics. Previously, Sanz-Forcada *et al.*[93] had found its X-ray counterpart with *XMM-Newton*, but were unable to determine its real nature. Later, other authors imaged σ Ori IRS1 with *Chandra*[13,78,80]. Finally, Hodapp *et al.*[94], using near-infrared integral-field spectroscopy, discovered that σ Ori IRS1 is actually a photo-eroded pair of a very-low-mass star and a brown-dwarf proplyd. The only astrometric observations of σ Ori IRS1 A and B have been made by Bouy *et al.*[41], who measured $\rho$ = 0.2363±0.024 arcsec, $\theta$ = 318.1±0.4 and $\Delta Ks$ = 2.19±0.01 mag on 2004.776 (with NACO/VLT), and $\rho$ = 0.2429±0.036 arcsec, $\theta$ = 317.0±0.7 and $\Delta Ks$ = 2.17±0.07 mag on 2007.914 (with MAD/VLT).

- Mayrit 11238A,B = σ Ori Ca,Cb ($N$ = 7; see Table I). The familiar star σ Ori C has been imaged virtually tens of thousands times with telescopes of all sizes, but has been poorly investigated spectroscopically[95]. Again, its companion candidate was discovered by Caballero[92], first measured by Caballero[48] ($\rho \sim$ 2.0 arcsec, $\theta \sim$ 20 deg on 2003.688), and characterised astrometrically in detail by Bouy *et al.*[41] ($\rho$ = 1.9915±0.0039 arcsec, $\theta$ = 11.5±0.7 deg on 2007.914). The magnitude differences between σ Ori Ca and Cb, of 5.50±0.07 mag in $H$ and 5.05±0.14 mag in $Ks$, would locate the companion in the very-low-mass-star domain, close to the substellar boundary, if it belonged to the cluster.

- Mayrit 13084 = σ Ori D ($N$ = 8; see Table I and Fig. 1). It is a bright, single, normal B2 V star with a faint (wind-driven?) X-ray emission[13,95].

- Mayrit 21023, 30241, and 36263 ($N$ = 11; see Table II). They are three cluster members with known features of youth[11] closer to the cluster centre than σ Ori E.

- Mayrit 42062 A,B = σ Ori Ea,Eb ($N$ = 13; see Table I). The famous, helium-rich, magnetically active, radio and X-ray emitter, short-period rotationally variable, spectroscopically peculiar star σ Ori E (B2 Vp; 7–8 $M_\odot$) has been extensively studied in the literature[95-107]. However, nobody has yet tried to confirm, or use in their rather complex models, the tight binarity claimed by Bouy *et al.*[41]. They found a companion candidate to σ Ori E, 3–4 mag fainter in *Ks*, at $\rho \sim$ 0.330 arcsec and $\theta \sim$ 301 deg on 2007.914 (note the corrected angle; Bouy, priv. comm.). According to Caballero *et al.*[13,108], the numerous X-ray flares[109] may come from the K–M-type secondary at 100–150 AU, while the X-ray low-amplitude modulation may have its origin in the plasma trapped in heterogeneous magnetospheric clouds that transit across the disc of the B2 Vp primary[110].

- Mayrit 50279, 53049 and 53144 ($N$ = 16; see Table II). They are another three young stars within the innermost 1 arcmin, but further to the cluster centre than σ Ori E. Interestingly, Mayrit 50279 is with $J \approx$ 14.0 mag and $M$ = 0.09–0.08 $M_\odot$ the faintest object with youth features in the area[11], even fainter than the brown-dwarf proplyd σ Ori IRS1 B.

I have not enumerated the cluster member candidate [BNL2005] 3.01 67 (at $\rho$ = 53.42±0.06 arcsec, $\theta$ = 144.16±0.02 deg to σ Ori A), which displays low-gravity features, manifested by weak pseudo-equivalent widths of alkali lines[111], and very faint X-ray emission[13,80,112], but which seems to be actually an M2.5 V spectroscopic binary in the field[26].

Within a 60 arcsec-radius circle, there are many more catalogued sources with unknown membership status[41]. However, most of them (including the relatively bright sources 2MASS J05384652–0235479 and 2MASS J05384454–0235324[77,78]), with only two- or one-band photometry, must be faint reddened stars and galaxies in the background under reasonable assumptions of spatial distribution, mass function, and contamination[30,113,114,115].

*Close binaries outside the cluster centre*

To date, apart from the σ Ori system itself, only another five bright stars in the cluster have been targeted with high-resolution (0.15–0.25 arcsec) imaging with adaptive optics (with NAOMI at the 4.2 m William Herschel Telescope at epoch 2003.688)[48,92]. Two of them, the stars Mayrit 306125 and 528005, turned out to be close binaries with angular separations of 0.4–0.5 arcsec. Another eight cluster pairs with angular separations between ~0.5 and ~3 arcsec (~200–1200 AU at the cluster distance) have been reported by the author in other works from data obtained with the 2.5 m Isaac Newton[48], 1.5 m Carlos Sánchez and 0.80 m IAC80 telescopes[48,78], asymmetries of stellar profiles in 2MASS images[11], and digitized photographic plates[13,116]. The ten close-binary candidates, which are not fully resolved into two in standard observations from the ground, are listed in Table III. Some pairs lack relevant data, such as accurate angular separation, position angle, or magnitude difference in a certain photometric band. However, this gap will soon be filled by an analysis of UKIDSS near-infrared data[117,118]. "NE" and "SE" in position angle stand for reported elongations in the north-east and south-east directions, respectively. All primaries, being either in physical or visual systems, are very young mid-K to mid-M dwarfs except for the also young B5 V star Mayrit 306125.

Table III
*Reported close binaries not listed in WDS and at ρ > 1.0 arcmin to σ Ori A*

| Star (Mayrit No.) | ρ [arcsec] | θ [deg] | Δmag (band) [mag] |
|---|---|---|---|
| [W96] rJ053847−0237 (92149 AB)[8] | ~1.9[78] (2.1)[30] | ~60[78] (64)[30] | 1.1±0.4 (V)[78] 0.80±0.17 (R)[78] 0.62±0.10 (I)[78] 0.337 (Z)[30] 0.381 (Y)[30] 0.468 (J)[30] 0.548 (H)[30] 0.828 (K)[30] |
| [W96] rJ053834−0234 (168291 AB)[26] | ~3.5[13] | NE[13] | ... |
| HD 37525 (306125 AB) | 0.47±0.04[92] | 189±7[92] | ≥0.5 (H)[92] |
| [W96] 4771−899 (528005 AB) | 0.40±0.08[92] | 170±10[92] | 0.337±0.019 (H)[92] |
| [W96] rJ053859−0247 (707162 AB) | ~1.0[48] | ... | ~0.0 (R)[48] |
| 2E 1486 (1106058 AB) | <3[11] | ... | ... |
| V605 Ori (1245057 AB) | <3[11] | ... | ... |
| [SWW2004] 118 (1411131 AB) | <3[11] | ... | ... |
| 2E 1464 (1564349 AB) | ~0.5[48] | ... | ~0.8 (R)[48] |
| [SE2004] 6 (1610344 AB) | ~3.0[116] | SE[116] | ... |

Of the ten close binary candidates, only one has been investigated as a double system in some detail: Mayrit 92149 AB. Caballero[78] proposed that the primary had a circumstellar disc from the magnitude differences from the blue optical to the near-infrared. Earlier, Sherry et al.[58] had tabulated two objects separated by 0.75 arcsec and with roughly the same $VRI_cJHK_s$ magnitudes in the coordinates of

Mayrit 92149 (SWW 102 and 149), but it is not known whether they actually recognized the pair (they also tabulated two objects, SWW 4 and SWW 159, separated by only 0.011 arcsec at the coordinates of the single star Mayrit 126250). As well, Lodieu *et al.*[30] recovered and tabulated accurate coordinates and magnitudes of the two stars in the previously-known pair using UKIDSS data[117], from where I derived the $\rho$, $\theta$ and $\Delta mag$ values in parenthesis in Table III.

*Wide binaries outside the cluster centre*

Some photometric surveys for new σ Orionis member candidates have identified additional young binary candidates with angular separations $\rho > 4$ arcsec that are not tabulated by the WDS and still require a careful study. Table IV lists the 16 published pairs in 12 systems, together with discovery references and recomputed $\rho$, $\theta$, and $\Delta K$ values from UKIDSS data as in Lodieu *et al.*[30]. The new angular separations, position angles and magnitude differences match previous determinations within uncertainties. The tabulated magnitude differences of the two companions of Mayrit 208324 are actually $\Delta K_s$ because of saturation of the primary in the UKIDSS image. The values of $\rho$, $\theta$, and $\Delta K$ of the two systems containing the 'isolated planetary-mass object' *candidates* S Ori 68 and S Ori 74 were derived from VISTA data[31].

I provide for the first time with Mayrit numbers for the cluster members and candidates S Ori 68 (483350), S Ori J053926.8–022614 (860047) and 2MASS J05393660–0242222 (866116). However, I do not provide with them for the interloper stars and distant galaxies S Ori 74, S Ori J053946.3–022631, HD 37686 #2, 2MASS J05381428–0210177, and USNO-A2.0 0825–01615246.

Following the author' cluster centre separation-pair separation diagram[119], only three pairs among the nine possible cluster binaries in Table IV would have probabilities of chance alignment < 1 %:

- Mayrit 487350 + 483350 ($\rho \sim 4.6$ arcsec). It is the well known brown dwarf–'planetary-mass object' pair candidate [SE2004] 70 + S Ori 68, which components still miss an irrefutable confirmation of true membership in σ Orionis[30,31,120].
- Mayrit 856047 + 860047 ($\rho \sim 4.6$ arcsec). It is a curious example of X-ray emission assigned originally to a faint brown dwarf candidate[121] that later was found to be the secondary of a close pair[12]. The high-energy emission is originated instead in the primary, which is a very low-mass star 2.8 mag brighter in the *J* band.
- Mayrit 863116 + 866116 ($\rho \sim 4.9$ arcsec). The primary is a T Tauri star that has been investigated spectroscopically[75,122,123] and that its X-ray emission was already detected by the *Einstein* space mission. The companion candidate has been catalogued by 2MASS and SDSS DR9[124], and its optical and near-infrared colours are consistent with cluster membership. Because of its relatively wide separation and brightness ($r' \sim 15.0$ mag), it is a suitable spectroscopic target for 4 m-class telescopes or larger. If confirmed, the presence of a nearby young M-type component may explain the peculiar X-ray variability of Mayrit 863116 (it displayed a flaring event over-imposed over a sinusoidal short-term variation[13,108]). Besides, there is a second, much fainter companion candidate to Mayrit 863116 at $\rho \sim 5$ arcsec to the NWN, which is probably in the background.

There are also three more systems with angular separations $\rho \sim$ 8–12 arcsec and low probabilities of chance alignment at ~1 %: Mayrit 528005 AB + 530005, Mayrit 968292 + 958292, and Mayrit 1415279 AB + 1416280. At least one of the components in each pair shows the Li I spectral line in absorption (the two components in the case of Mayrit 528005 AB + 530005), while two of the three "pairs" may actually be triple because of spectroscopic binarity of one of the components.

Table IV
*Wide binaries not listed in WDS and at $\rho$ > 1.0 arcmin to σ Ori A*

| Primary (Mayrit No.) | Secondary (Mayrit No) | $\rho$ [arcsec] | $\theta$ [deg] | $\Delta K$ [mag] |
|---|---|---|---|---|
| HD 294271 (208324) | [HHM2007] 614 (214321)[48,79,92] | 10.62±0.02 | 266.7±0.2 | (4.63±0.06) |
|  | [HHM2007] 606 (219320)[48,79,92] | 18.72±0.02 | 268.8±0.2 | 4.83±0.05 |
| [W96] 4771–1051 (260182) | S Ori J053844.4–024030 (270181)[29,48] | 11.46±0.02 | 160.3±0.2 | 2.0856±0.0016 |
|  | S Ori 74 (...)[125] | 11.90±0.10 | 3.2±1.0 | 9.0: |
|  | S Ori J053844.4–024037 (277181)[29,48] | 18.34±0.02 | 168.1±0.2 | 3.477±0.004 |
| V507 Ori (397060) | S Ori 7 (410059)[48,126] | 14.10±0.02 | 40.6±0.2 | 2.638±0.002 |
| [SE2004] 70 (487350) | S Ori 68 (483350)[12,120] | 4.59±0.10 | 136.0±1.0 | 4.0±0.3 |
| [W96] 4771–899 (528005 AB) | S Ori J053847.5–022711 (530005)[48,92] | 7.69±0.02 | 286.9±0.2 | 1.5934±0.0008 |
| [SE2004] 94 (856047) | S Ori J053926.8–022614 (860047)[120,121] | 4.64±0.02 | 75.6±0.2 | 2.802±0.015 |
| 2E 1484 (863116) | 2MASS J05393660–0242222 (866116)[127] | 4.92±0.02 | 167.9±0.2 | 1.6360±0.0007 |
| [W96] 4771–962 (968292) | SO210868 (958292)[48,68] | 9.92±0.02 | 97.4±0.2 | 0.3869±0.0007 |
| V2750 Ori (1087058) | S Ori J053946.3–022631 (...)[57,128] | 5.87±0.02 | 265.4±0.3 | ≥6.0 |
| HD 37686 (1359007) | HD 37686 #2 (...)[48] | 5.50±0.02 | 203.8±0.3 | 5.68±0.02 |
| OriNTT 429 (1415279 AB) | [SWW2004] 22 (1416280)[126,129] | 12.15±0.02 | 4.2±0.2 | 2.5683±0.0019 |
| [SE2004] 6 (1610344) | 2MASS J05381428–0210177 (...)[116] | 4.85±0.02 | 240.2±0.3 | 4.25±0.02 |
|  | USNO-A2.0 0825–01615246 (...)[116] | 8.18±0.02 | 244.1±0.3 | 6.8±0.2 |

*Spectroscopic binaries*

In spite of the attention given to radial-velocity surveys for spectroscopic binaries in σ Orionis, between the first report of a suspected spectroscopic binary in 1904 (of σ Ori A itself[87]) and the first confirmed one with measurement of lithium abundance in its two components[130], a century had to pass. In its turn, between the discovery of this K-type, weak-line T Tauri, double-line spectroscopic binary and the first dedicated radial-velocity surveys, another decade elapsed[25,26,112] (but see Wolk[8]). Regardless of having used large facilities (e.g., AF2/WYFFOS at the William Herschel Telescope, FLAMES at the Very Large Telescope), only 16 spectroscopic binary candidates are known to date, and only four of them have accurate orbit determinations[26,40].

I compile in Table V the 16 known reliable single-line (SB1) and double-line (SB2) spectroscopic binaries in σ Orionis. They are not the only SB candidates reported in the cluster area. For example, Kenyon *et al.*[25] found binarity of the field star [BZR99] S Ori 20[116]. Besides, Maxted *et al.*[112] claimed binarity of three photometric cluster member candidates, one of which was later classified as a non-member SB1[26]; the membership status of the other two stars remain unknown, as no photometric survey has picked them up. The bright stars Mayrit 1116300 (HD 37333, unresolved with adaptive optics[48,92]), Mayrit 524060 (HD 37564, for which previous studies have suggested that is probably not a member of the Orion OB1 association[131]), and σ Ori B have also been proposed to be SBs, but solely based on wide-band photometry[132,133]. Finally, according to Alcalá *et al.*[122,123], the high-resolution spectra of Mayrit 863116 (2E 1484) displays a broad cross-correlation function, probably due to its high rotational velocity of $v\sin i \approx 150$ km s$^{-1}$, and not to unresolved spectroscopic binarity[75]. In Table V, I also provide for the first time with a Mayrit number for the cluster member S Ori 36.

Table V
*Known spectroscopic binaries in σ Orionis*

| Mayrit No. | Binary | Type | P [d] |
|---|---|---|---|
| Aab | σ Ori Aa,Ab | SB2[40] | 143.5±0.5[40] |
| 102101 AB | [W96] rJ053851–0236 | SB2[26] | 8.72±0.02[26] |
| 114305 AB | [W96] 4771–1147 | SB2[8] | ... |
| 240355 AB | [SWW2004] 144 | SB2[26] | 8.52±0.01[26] |
| 258337 AB | [HHM2007] 633 | SB1[112] | ... |
| 332168 AB | [SWW2004] 205 | SB1[26] | ... |
| 344337 AB | 2E 1468 | SB1[26] | ... |
| 453037 AB | [W96] rJ053902–0229 | SB2[8] | ... |
| 459224 AB | S Ori J053823.6–024132 | SB2[112] | ... |
| 547270 AB | Kiso A–0976 316 | SB1[26] | ... |
| 633095 AB | S Ori 36 | SB1[25] | ... |
| 873229 AB | Haro 5–7 | SB2[112] | ... |
| 1415279 AB | OriNTT 429 | SB2[130] | 7.49[130] |
| 1436317 AB | [KJN2005] 72 | SB1[25] | ... |
| 1482130 AB | S Ori J054000.2–025159 | SB1[25] | ... |
| 1493050 AB | [OJV2004] 24 | SB2[112] | ... |

*Triples (and higher orders)*

I conclude this review on multiplicity in σ Orionis with an enumeration of the triple (and sextuple) system candidates in the cluster, apart from the massive triple σ Ori Aa, Ab, B.

- Mayrit 528005 AB + 530005 ([W96] 4771–899 AB–C). The system is made up of a bright K3 binary T Tauri star of $s \sim 140$ AU[8,92] (Table III) and a mid-M candidate companion with youth features at $s \sim 3000$ AU[23,79] (Table IV), and with a probability of chance alignment of only ~1 %.

- Mayrit 1415279 AB + 1416280 (OriNTT 429 AB–C). It is a system very similar to the previous one, except for the primary being a K2–3 spectroscopic binary with $P = 7.94$ d[48,130] (Table V) and the secondary being a photometric cluster member at $s \sim 4700$ AU[11,58] (Table IV ) in the outskirts of the cluster.

- Mayrit 260182 + 270181 + 277181 ([W96] 4771–1051 A–B–C). This curious triple asterism is formed by two low-mass stars and a high-mass brown dwarf, all of them with youth features, separated by 4400–7100 AU[23] (Table IV). Bihain *et al.*[125] reported a fourth component candidate with a substellar mass below the deuterium burning limit, namely S Ori 74, but new, more accurate VISTA photometric data do not support the candidacy of this object[31].

- Mayrit 208324 + 214321 + 219320 + 182305 + 189303 + 240322 (BD–02 1342A–F). Struve[50] catalogued the quadruple system STF 761 (WDS 05386–0233) for the first time (Table I). It consists of four bright early-type stars: HD 294271 [A], HD 294272 [BC], which is in turn a binary separated by $\rho = 8.54$ arcsec[49] ($s \sim 3300$ AU), and the poorly investigated star BD–02 1342D [D] (B5 V, B9.5 III, B8 V, and probably late A, respectively[75,134,135]). Two centuries later, two much fainter companions were discovered at $s \sim$ 4100–7200 AU to HD 294271[48,92], which have been reported to be bona fide cluster members in subsequent works[26,31,79] (Table IV). Therefore, WDS 05386–0233 becomes the only known sextuple system in the σ Orionis cluster.

*Summary and conclusions*

I made a comprehensive literature review on multiplicity in σ Orionis. First, I enumerated 16 stars in the innermost arcminute in the cluster, which are candidate members in the σ Ori trapezium. All of them but Mayrit 11238 B (σ Ori Cb) are confirmed young cluster members in a deep gravitational well of about 60 $M_\odot$. If they were gravitationally bound, σ Ori would be a sexdecuple system, which would increase to an *N*-tuple with $N > 16$ if any of the numerous faint photometric cluster member candidates reported by Bouy *et al.*[41] turned out to show youth features. But where do the multiple system finish and the open cluster begin? Actually, does this question have a meaning now, if in less than 10 Myr the system will be torn apart because of a pair of supernova explosions (σ Ori Aa, Ab), some main-sequence turn-offs (D, E, and, perhaps, B), and a consequent dramatic mass drop? These questions are connected to the Mayrit nomenclature, introduced by the author[78], which does not follow the classical terminology to naming the bright stars in the trapezium, but instead equalizes the cluster and the multiple system (e.g., σ Ori C, which is located at $\rho \approx 11$ arcsec and $\theta \approx 238$ deg with respect to the cluster centre defined by the bottom of the gravitational well, is Mayrit 11238).

Outside the central arcminute, only ten close binaries with angular separations in the approximate interval 0.4–3.0 arcsec have been reported (by the author). Of them, only two have been imaged with adaptive optics, which shows a clear lack of high-resolution imaging surveys for close binaries in the σ Orionis cluster.

I also listed other reported systems with $\rho > 4.0$ arcsec. Of the 16 tabulated systems, 11 have both primaries and secondaries that pass cluster-membership criteria, but only six (three) have probabilities of alignment of the order of 1% (less than 1%). It is still under debate whether these six wide systems are physically bound today or will remain so until the eventual disruption of the σ Orionis cluster within the Galactic disc.

Of the 16 known spectroscopic binaries in σ Orionis, nine are double-line binaries and only four have been determined their orbital periods, which range between 7–9 d for the low-mass pairs and ~140 d for the eponymous σ Ori Aa, Ab, B triple system. Finally, I report one sextuple system and three triples, of which one involves a spectroscopic binary and one a tight binary resolved only with adaptive optics.

As mentioned in the introduction, this review on multiplicity is limited by the heterogeneity and incompleteness of the data. Any attempt of deriving a frequency of multiplicity of the cluster from these data, of about 10% if one accounts for the kown cluster members and multiple systems compiled here, has no sense or, at least, must be understood only as an attempt of imposing a lower limit on that frequency. However, this review is useful for guiding future multiplicity studies in σ Orionis. Next steps can be, for example: (*i*) High-resolution imaging surveys with public UKIDSS and VISTA data down to $\rho \sim 0.4$ arcsec and/or with inexpensive lucky imagers at 2 m-class telescopes down to $\rho \sim 0.2$ arcsec. Adaptive-optics systems at larger telescopes could help with the follow-up and characterization of the most interesting companion candidates (*e.g.*, below the hydrogen burning limit). (*ii*) High-resolution spectroscopic surveys with spectrographs at 4-8 m-class telescopes for measuring accurately orbital periods of known single- and double-line spectroscopic binaries, for mass determination, and look for new ones. (*iii*) Intermediate-resolution spectroscopy for characterizing some poorly-investigated secondaries, with questionable youth features or even no spectral-type determination at all.

Unfortunately, a radial-velocity survey at the 10 m s$^{-1}$ level for disentangling between true and false pairs within the cluster, which would be the definitive

"binary test" of wide pairs, is out of the capabilities of current technology.

*Acknowledgements*

I thank Brian D. Mason for his kind comments and thoughts. Financial support was provided by the Spanish MICINN/MINECO under grants RYC2009-04666 and AYA2011-30147-C03-03. I thank I. Novalbos, T. Tobal and F. X. Miret for conceiving the idea of looking for binaries in σ Orionis, which eventually led to this work.

*References*

(1) R. H. Allen, *Star Names – Their Lore and Meaning*, ed. G. E. Stechert, New York, 1899.
(2) F. Verbunt & R. H. van Gent, *A&A*, **544**, A31, 2012.
(3) T. Brahe 1598, *Stellarum octavi orbis inerrantium accurata restitutio*, 1598 (re-ed. J. L. E. Dreyer, Copenhagen, 1916).
(4) J. Bayer, *Uranometria: omnium asterismorum continens schemata, nova methodo delineata, æreis laminis expressa*, ed. C. Magnus, Augsburg, 1603.
(5) J. Flamsteed, *Historia Cœlestis Britannica*, eds. I. Newton & E. Halley, London, 1712.
(6) R. F. Garrison, R. F., *PASP*, **79**, 433, 1967.
(7) G. Lyngå, *The Catalogue of Open Star Clusters, ADCBu*, **1**, 90, 1981.
(8) S. J. Wolk, S. J., Ph.D. thesis, State University of New York, 1996.
(9) V. J. S. Béjar, M. R. Zapatero Osorio & R. Rebolo, *ApJ*, **521**, 671, 1999.
(10) F. M. Walter *et al., Handbook of Star Forming Regions*, **1**, 732, 2008.
(11) J. A. Caballero, *A&A*, **487**, 667, 2008c.
(12) E. Franciosini, R. Pallavicini & J. Sanz-Forcada, *A&A*, **446**, 501, 2006.
(13) J. A. Caballero, J. F. Albacete-Colombo & J. López-Santiago, *A&A*, **521**, A45, 2010.
(14) J. M. Oliveira *et al., MNRAS*, **369**, 272, 2006.
(15) M. R. Zapatero Osorio *et al., A&A*, **472**, L9, 2007.
(16) K. L. Luhman *et al., ApJ*, **688**, 362, 2008.
(17) B. Reipurth *et al., Nature*, **396**, 343, 1998.
(18) S. M. Andrews *et al., ApJ*, **606**, 353, 2004.
(19) C. A. L. Bailer-Jones & R. Mundt, *A&A*, **367**, 218, 2001.
(20) J. A. Caballero *et al., A&A*, **424**, 857, 2004.
(21) A. Scholz *et al., MNRAS*, **398**, 873, 2009.
(22) A. M. Cody & L. A. Hillenbrand, *ApJ*, **741**, 9, 2011.
(23) M. R. Zapatero Osorio *et al., A&A*, **384**, 937, 2002.
(24) D. Barrado y Navascués *et al., A&A*, **393**, L85, 2002.
(25) M. J. Kenyon *et al., MNRAS*, **356**, 89, 2005.
(26) G. G. Sacco *et al., A&A*, **488**, 167, 2008.
(27) M. R. Zapatero Osorio *et al., Science*, **290**, 103, 2000.
(28) V. J. S. Béjar *et al., ApJ*, **556**, 830, 2001.
(29) J. A. Caballero *et al., A&A*, **470**, 903, 2007.
(30) N. Lodieu *et al., A&A*, **505**, 1115, 2009.
(31) K. Peña Ramírez *et al., ApJ*, **754**, 30, 2012.
(32) C. Mayer, *De novis in cœlo sidereo phænomenis in miris stellarum fixarum comitibus*, ed. J. E. Bode, Mannheim, 1779.

(33) J. S. Schlimmer, *Journal of Double Star Observations*, **3**, 151, 2007.
(34) W. Herschel, *Philosophical Transactions of the Royal Society*, **72**, 112, 1782.
(35) W. H. Smyth, *A cycle of Celestial Objects*, ed. J. W. Parker, London, 1844.
(36) W. R. Dawes, *MmRAS*, **8**, 61, 1835.
(37) S. W. Burnham, *AN*, **130**, 257, 1892.
(38) W. D. Heintz, *AJ*, **79**, 397, 1974.
(39) J. A. Caballero, *MNRAS*, **383**, 750, 2008b.
(40) S. Simón-Díaz *et al., ApJ*, **742**, 55, 2011.
(41) H. Bouy *et al., A&A*, **493**, 931, 2009.
(42) J. A. Caballero, *El observador de estrellas dobles*, **11**, 108, 2013a.
(43) B. Mason *et al., AJ*, **122**, 3466, 2001.
(44) J. A. Docobo *et al., RMxAA*, **43**, 141, 2007.
(45) M. F. Skrutskie *et al., AJ*, **131**, 1163, 2006.
(46) F. van Leeuwen, *A&A*, **474**, 653, 2007.
(47) E. Høg *et al., A&A*, **355**, L27, 2000.
(48) J. A. Caballero, PhD thesis, Universidad de La Laguna, 2006 (summary in English at: J. A. Caballero, *Highlights of Spanish Astrophysics*, **5**, 79, 2010b).
(49) B. Mason *et al., AJ*, **143**, 124, 2012.
(50) F. G. W. Struve, *AN*, **14**, 249, 1837.
(51) F. G. W. Struve *et al., Dun Echt Observatory Publications*, **1**, 1, 1876.
(52) N. H. Turner *et al., AJ*, **136**, 554, 2008.
(53) J. F. W. Herschel & J. South, *Philosophical Transactions of the Royal Society*, **114**, 83, 1824.
(54) L. Plaut *et al., Bulletin of the Astronomical Institutes of the Netherlands*, **7**, 181, 1934.
(55) F. Damm, priv. comm. to WDS.
(56) W. B. Weaver & A. Babcock, *PASP*, **116**, 1035, 2004.
(57) V. J. S. Béjar *et al., AN*, **325**, 705, 2004.
(58) W. H. Sherry *et al., AJ*, **128**, 2316, 2004.
(59) http://www.oneminuteastronomer.com.
(60) A. Abergel *et al., A&A*, **410**, 577, 2003.
(61) D. Ward-Thompson *et al., MNRAS*, **369**, 1201, 2006.
(62) M. Compiègne *et al., A&A*, **447**, 205, 2007.
(63) B. P. Bowler *et al., AJ*, **137**, 3685, 2009.
(64) K. Ogura & K. Sugitani, *PASA*, **15**, 91, 1998.
(65) M. W. Pound, B. Reipurth & J. Bally, *AJ*, **125**, 2108, 2003.
(66) J. Pety *et al., A&A*, **435**, 885, 2005.
(67) J. Goicoechea *et al., A&A*, **456**, 565, 2006.
(68) J. A. Caballero *et al., A&A*, **491**, 515, 2008.
(69) B. B. Ochsendorf *et al., A&A*, in press, 2014 (arXiv:1401.7185).
(70) P. Padoan & Å. Nordlund, *ApJ*, **576**, 870, 2002.
(71) M. R. Bate *et al., MNRAS*, **339**, 577, 2003.
(72) S. P. Goodwin *et al., A&A*, **414**, 633, 2004.
(73) A. P. Whitworth & H. Zinnecker, *A&A*, **427**, 299, 2004.
(74) A. P. Whitworth *et al., Protostars and Planets*, **5**, 459, 2007.
(75) J. A. Caballero, *A&A*, **466**, 917, 2007a.
(76) J. A. Caballero, *The Star Formation Newsletter*, **243**, 6, 2013b.
(77) W. H. Sherry, F. M. Walter, S. J. Wolk, *Star Formation in the Era of Three Great Observatories*, 53, 2005.

(78) J. A. Caballero, *AN*, **328**, 917, 2007b.
(79) J. Hernández *et al., ApJ*, **662**, 1067, 2007.
(80) S. L. Skinner *et al., ApJ*, **683**, 796, 2008.
(81) M. L. Siegrist, *Vrania*, 227, 1951 (*Seminario de Astronomía y Geodesia de la Universidad de Madrid*, publicación no. 8).
(82) W. Kümmritz, in *IAU Commission 26 Information Circular* No. 14, 1958.
(83) W. D. Heintz, *AJ*, **79**, 819, 1974.
(84) W. I. Hartkopf, B. D. Mason & H. A. McAlister, *AJ*, **111**, 370, 1996.
(85) W. D. Heintz, *ApJS*, **111**, 335, 1997.
(86) N. H. Turner *et al., AJ*, **136**, 554, 2008.
(87) E. B. Frost & W. S. Adams, *ApJ*, **19**, 151, 1904.
(88) C. T. Bolton, *ApJ*, **192**, L7, 1974.
(89) C. Hummel *et al.*, oral contribution in *Massive stars: from α to Ω*, 10–14 Jun 2013, Rhodes, Greece, 2013.
(90) G. Schaefer *et al.*, oral contribution in *CHARA/NPOI collaboration meeting*, 18–20 Mar 2013, Flagstaff AZ, USA, 2013.
(91) J. T. van Loon & J. M. Oliveira, *A&A*, **405**, L33, 2003.
(92) J. A. Caballero, *AN*, **326**, 1007, 2005.
(93) J. Sanz-Forcada, E. Franciosini & R. Pallavicini, *A&A*, **421**, 715, 2004.
(94) K. W. Hodapp *et al., ApJ*, **701**, L100, 2009.
(95) J. L. Greenstein & G. Wallerstein, *ApJ*, **127**, 237, 1958.
(96) D. A. Klinglesmith *et al., ApJ*, **159**, 513, 1970.
(97) H. Pedersen & B. Thomsen, *A&AS*, **30**, 11, 1977.
(98) J. D. Landstreet & E. F. Borra, *ApJ*, **224**, L5, 1978.
(99) D. Groote & K. Hunger, *A&A*, **116**, 64, 1982.
(100) S. A. Drake *et al., ApJ*, **322**, 902, 1987.
(101) F. Leone & G. Umana, *A&A*, **268**, 667, 1993.
(102) D. Groote & K. Hunger, *A&A*, **319**, 250, 1997.
(103) A. Reiners *et al., A&A*, **363**, 585, 2000.
(104) R. H. D. Townsend, S. P. Owocki & D. Groote, *ApJ*, **630**, L81, 2005.
(105) R. H. D. Townsend *et al., ApJ*, **714**, L318, 2010.
(106) M . E. Oksala *et al, MNRAS*, **419**, 959, 2012.
(107) A. C. Carciofi *et al., ApJ*, **766**, L9, 2013.
(108) J. A. Caballero *et al., AJ*, **137**, 5012, 2009.
(109) D. Groote & J. H. M. M. Hunger, *A&A*, **418**, 235, 2004.
(110) R. H. D. Townsend *et al., ApJ*, **769**, 33, 2013.
(111) B. Burningham *et al., MNRAS*, **356**, 1583, 2005.
(112) P. F. L. Maxted *et al., MNRAS*, **385**, 2210, 2008.
(113) J. A. Caballero, *MNRAS*, **383**, 375, 2008a.
(114) J. A. Caballero, A. J. Burgasser & R. Klement, *A&A*, **488**, 181, 2008.
(115) V. J. S. Béjar *et al., ApJ*, **743**, 64, 2011.
(116) J. A. Caballero, *A&A*, **514**, A18, 2010a.
(117) A. Lawrence *et al., MNRAS*, **379**, 1599, 2007.
(118) J. A. Caballero *et al., A&A*, subm.
(119) J. A. Caballero, *Highlights of Spanish Astrophysics*, **5**, 377, 2010c.
(120) J. A. Caballero *et al., A&A*, **460**, 635, 2006.
(121) F. Mokler & B. Stelzer, *A&A*, **391**, 1025, 2002.
(122) J. M. Alcalá *et al., A&AS*, **119**, 7, 1996.
(123) J. M. Alcalá *et al., A&A*, **353**, 186, 2000.

(124) C. P. Ahn *et al., ApJS*, **203**, 21, 2012.
(125) G. Bihain *et al., A&A*, **506**, 1169, 2009.
(126) J. A. Caballero, *Proceedings of the 15th Cambridge Workshop on Cool Stars, Stellar Systems and the Sun*, AIPC, **1094**, 912, 2009b.
(127) I. Novalbos, T. Tobal & F. X. Miret, priv. comm.
(128) B. M. González-García *et al., A&A*, **460**, 799, 2006.
(129) J. López-Santiago & J. A. Caballero, *A&A*, **491**, 961, 2008.
(130) C. W. Lee, C. W., E. L. Martín & R. D. Mathieu, *AJ*, **108**, 1445, 1994.
(131) A. G. A. Brown, E. J. de Geus & P. T. de Zeeuw, *A&A*, **289**, 101, 1994.
(132) H. M. Tovmassian, R. Kh. Hovhannessian & R. A. Epremian, *Astrofizika*, **35**, 5, 1991.
(133) W. H. Sherry *et al., AJ*, **135**, 1616, 2008.
(134) R. E. Schild & F. Chaffee, *ApJ*, **169**, 529, 1971.
(135) H. H. Guetter, *AJ*, **86**, 1057, 1981.

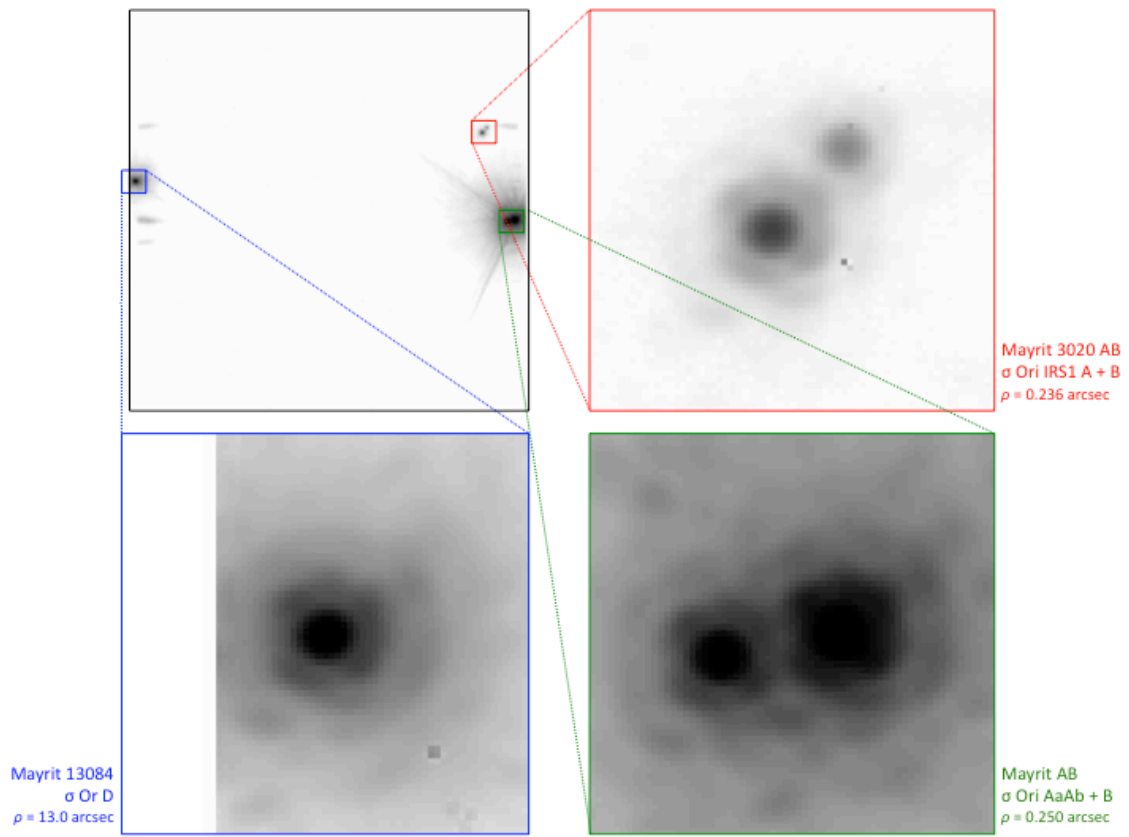

FIG. 1

Colour-inverted *Ks*-band image of σ Ori Aab, B, IRS1A, IRS1B and D taken with NACO at the Very Large Telescope UT4 in October 2004 during instrument commissioning for astrometric calibration. *Top left panel*: full field of view of approximate size of 13.6 × 13.6 arcsec$^2$. *Remaining panels*: sub-images, about 1 arcsec in size, centred on σ Ori IRS1 A,B (Mayrit 3020 AB; *top right*), σ Ori Aab,B (Mayrit AB; bottom right), and σ Ori D (Mayrit 13084), at 13.0 arcsec to the east of σ Ori Aab,B. North is up, east is to the left. Names and angular separations are labeled.